\documentclass[aps,pra,groupedaddress,showpacs,showkeys]{revtex4}

\usepackage{amsmath}
\usepackage{graphicx}

\begin{document}

\title{An easy method to estimate charge-exchange cross sections 
between ions and one-active-electron atoms}

\author{Fabio Sattin }

\email{sattin@igi.pd.cnr.it}

\affiliation{Consorzio RFX, Associazione Euratom-ENEA , Corso Stati Uniti 4, 
35127 Padova, Italy}

\begin{abstract}
In this paper we present a simple model for computing single 
electron exchange between light nuclei. The electronic capture  
is pictured as a tunnelling process in a model 
potential. This allows to analytically compute the transmission 
probability. Some comparisons with data from existing literature are given.
\end{abstract}
 


\date{\today}

\maketitle

\section{Introduction}
Charge exchange processes between atomic particles are of great 
importance in plasma physics and astrophysics. By example, it is 
through this mechanism that energetic charged particles can escape 
from the core of magnetic confinement devices; conversely, cold neutral particles 
coming from the wall can diffuse towards the centre. \\
While only quantum mechanical methods can give really accurate 
computations of all of the basic quantities for these processes, i.e. 
total and partial or differential cross 
sections, less precise but simpler methods can still be useful. 
In some cases they can still give high-precision estimates. 
It is the case of the Classical Trajectory Monte 
Carlo (CTMC) method, successfully applied to high-velocity 
collisions (but recently extended also to the low velocity range, see, e.g.,
\cite{rakovic,schultz} for a discussion and some recent improvements on 
this subject), to Rydberg-atom collisions, and to multiply-charged 
nuclei. 
In the low-velocity region, analytical or semi-analytical methods are 
often used: we mention here just the Over Barrier Models (OBMs). They are known 
since a long time \cite{ryufuku} and are still being improved to 
include as much physics as possible 
\cite{niehaus,ostrovsky,jpb,pra00,pra01}. \\
It is worthwhile to notice that, although the computation of 
charge exchange cross sections is a well developed field of research 
since several decades, its techniques are by no means completely established: 
it may still happen that theoretical 
predictions be confuted by experiments or that different methods give 
sharp discrepancies even in relatively simple situations, that should 
be quite well diagnosed  by now \cite{caillar}. Therefore, the 
development of new, different methods of computation can still be valuable.
In this work we suggest a fast algorithm to 
estimate single electron captures. From the tests that we show in this 
paper, it appears also rather accurate. It is rooted upon standard OBMs, 
in that the electronic capture process is regarded as a potential barrier 
crossing. Unlike OBMs, however, the electron is considered as a 
quantum-mechanical object. This allows to compute also under-barrier 
crossing events (tunnelling). In order to reduce the problem to 
a manageable, semi-analytical form, several geometrical 
simplifications will be done. The results are compared against 
experimental data as well as other different theoretical computations, 
and are found to fit them nicely. 

\section{Description of the model}
We consider a scattering experiment between a nucleus {\bf T} with one
active electron {\bf e}, and a projectile nucleus {\bf P}. Let $\rho$ 
be the electron position relative to {\bf T} and $R$ the relative 
distance between {\bf T} and {\bf P} (see Fig. 1). Several approximations are 
necessary:
I) All the nuclei are regarded as hydrogenlike particles,
thus $Z_{p}$ and $Z_{t}$ are the effective charge of the projectile and  
of the target seen by the electron, respectively;
II) The two nuclei are considered 
as approaching slowly if compared to the orbital electron velocity
(adiabatic approximation). This means that the electron is allowed to 
complete its path from the target to the projectile before 
any appreciable relative movement of the nuclei occurs. 
III) We adopt a straight-line approximation for 
the nuclear trajectory.  
IV) We neglect the 
possibility of target or projectile ionization. V) Finally, we discard 
also the possibility of electronic re-capture from the projectile by 
the target (although this possibility can be implemented within the 
algorithm without much effort). Points II) and III) are not mutually 
contradictory provided that impact velocity $u$ is not too small. 
We remind that the high- and low-velocity ranges are discriminated by $ 
u \ge v_{e}, u \le v_{e}$, 
with $v_{e}$ classical velocity of the electron. Point 
IV), too, is consistent only with slow collisions. Point V), finally, is 
likely to be more and more satisfied as the ratio $Z_{p}/Z_{t}$ 
increases well beyond the unity.\\ 

\begin{figure}
\includegraphics[scale=0.6]{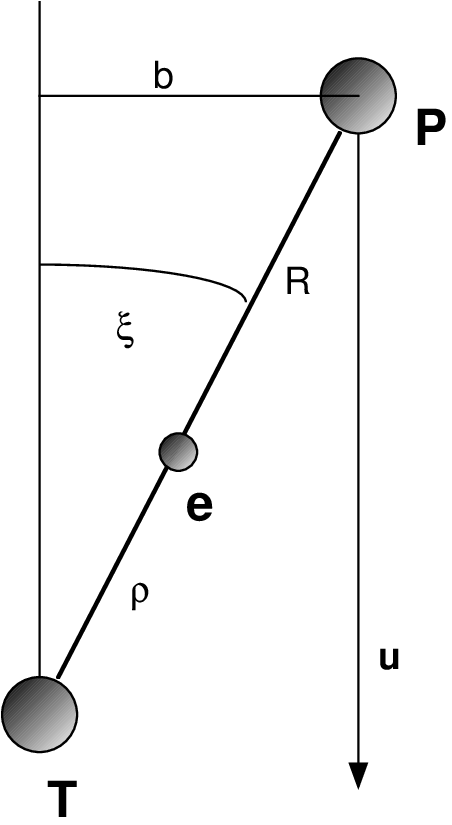}
\caption{Sketch of the scattering geometry}
\end{figure}

Let us for a while look at the electron as it were a classical 
particle, and assume that it is initially in a low angular momentum state 
(e.g. an $s$ state):  this means that its trajectory is a straight segment 
along the radial direction, starting from the target nucleus (Fig. 
1). \\
The total energy of the electron is
\begin{equation}
		\label{eq:energia}
E = {p^2 \over 2 } + U = {p^2 \over 2} - {Z_{t} \over \rho} - 
{Z_{t} \over R - \rho}
\end{equation}
(Atomic units will be used unless otherwise stated). 
It is straightforward to work out the value and the position of the 
maximum of the potential $U$ along the internuclear axis, as depicted in Fig. 2:
\begin{equation}
\label{eq:um}		
	U_{M}(\rho_{M}) = - {1 \over R} \left(\sqrt{Z_{t}} + 
	\sqrt{Z_{p}}\right)^{2} \quad ,
\end{equation}
\begin{equation}
		\rho_{M} = R{ \sqrt{Z_{t}} \over \sqrt{Z_{t}} + \sqrt{Z_{p}}} \quad .
\end{equation}
\begin{figure}
	\includegraphics[scale=0.6]{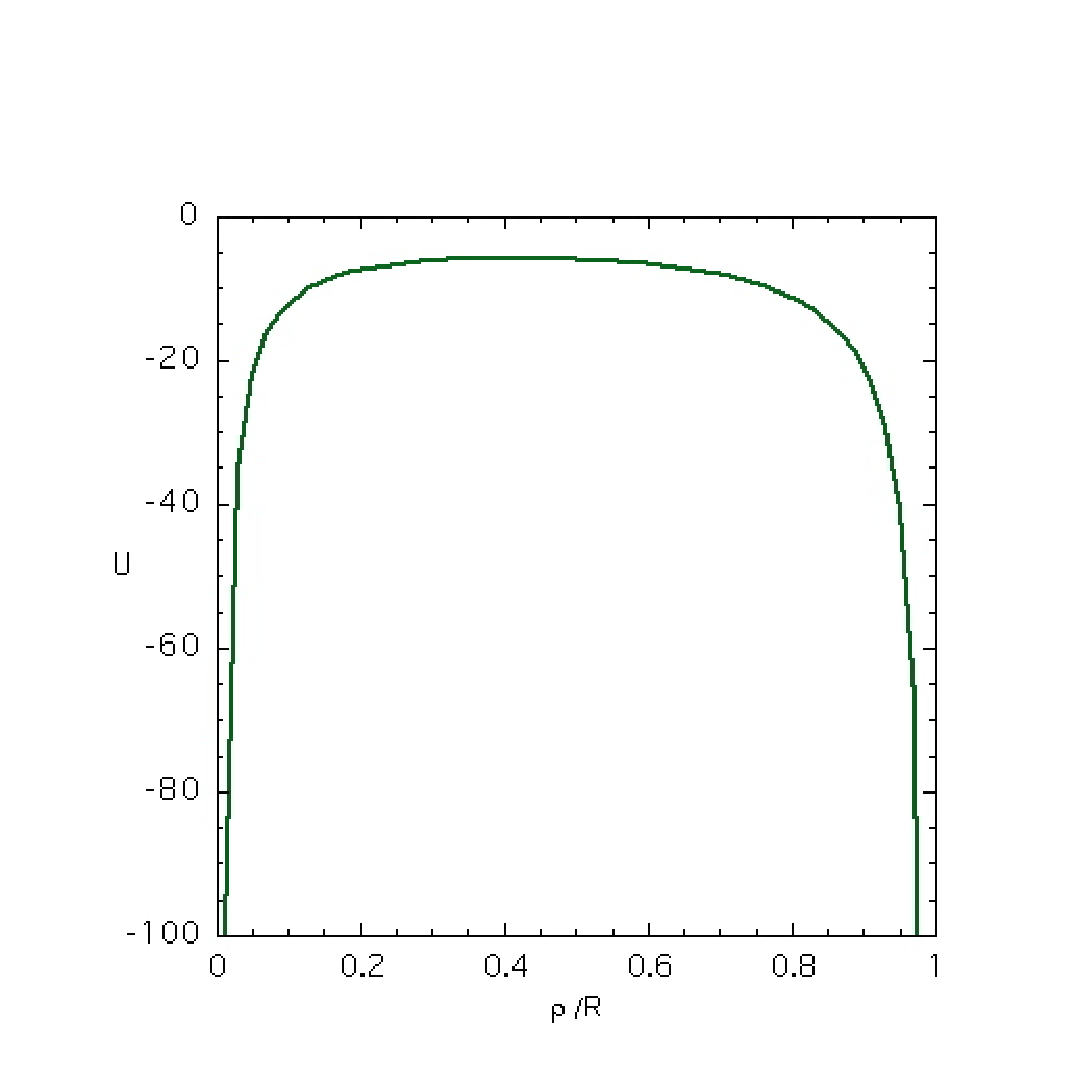}
	\caption{\label{fig:U} Potential $U$ along the internuclear axis.}
\end{figure}
We make the hypothesis that only nearly head-on collisions of the 
electron on the projectile lead to capture: that is, captures occur only along or close to the internuclear 
axis. Let us define $p_{c}$ as the probability for an electron impinging 
exactly along this axis to be captured, and write the capture probability $f$ for 
electrons impinging within the solid angle $d\Omega = \sin(\xi) d\xi 
d\varphi $ around the internuclear axis as 
\begin{equation}
		f = p_{c} \delta(\varphi) { d\Omega \over 4 \pi} \quad ,
\end{equation}		
where $\delta$ is the Dirac delta and $ 4 \pi$ is a normalization 
factor accounting for an isotropic distribution of the electronic 
velocities. The presence of the delta function means that only electrons whose trajectories are 
completely lying in the same plane as the three particles are allowed to be captured. 

Because of the straight-line trajectory there is a one-to-one 
correspondence between time and angle $\xi$:
\begin{equation} 
		\xi = \arcsin\left({b \over R(t)}\right) \quad ,
\end{equation}
\begin{equation}
		R(t) = \sqrt{b^{2} + (u t)^{2}} \quad ,
\end{equation}
where $b$ is the impact parameter.\\
Let us now define $W$ as the probability for the electron to be 
still bound to {\bf T} at time $t$. Its rate of change as
given by
\begin{equation}
\label{eq:rate0}
dW(t)  =  - f W(t) = - W(t) p_{c} \delta(\varphi) { d\Omega \over 4 \pi} \quad . 
\end{equation}
The integration over the azimuthal angle $\varphi$ is 
straightforward, and we find
\begin{equation}
\label{eq:rate}
dW(t)  = -  W(t) {p_{c} \over 4 \pi } \sin(\xi) { d\xi (t) \over dt} 
dt \, 
dt  \quad .
\end{equation}
Integration over time, with the boundary condition $W(-\infty) = 1$, 
yields
\begin{equation}
\label{eq:w}
W(t) = \exp\left( - {1 \over 4 \pi } \int_{-\infty}^{t} p_{c}(\tau) 
\sin(\xi(\tau)) { d\xi(\tau) \over 
d\tau} \, d\tau \right)
\end{equation}
and we have put into evidence that the 
factor $p_{c}$ is a function of distance $R$ and thus of time. \\
The total capture probability is $ 
P(b) =  1 - W(\infty) $ and the total 
cross section is given by the integral of this quantity in the 
impact parameter space: $\sigma = 2 \pi \int b P(b) db$. \\
By reducing everything to one-dimensional geometry, we have chosen to 
place all of the important physics in the transmission factor $p_{c}$: the 
probability for an electron to cross the potential barrier. This is at a 
difference with, e.g., OBMs, where the spatial form of the potential 
is of critical importance (see \cite{ostrovsky,jpb,pra00}). \\
If we chose to maintain the classical picture for the electron, we could 
recover a very simplified version of Over Barrier Model, by putting 
$p_{c} = 1$ in the region classically allowed to the electron, and zero 
elsewhere. We instead choose to compute $p_{c}$ through a quantum-mechanical 
picture:  we model the process of transferring the electron from one 
nucleus to the other as a tunnelling process through the potential 
barrier, with the factor $p_{c}$ which becomes the transmission factor. 
Even for this simplified problem the quantum mechanical transmission factor 
can be computed only by complicated numerical techniques. Since the goal of this 
paper is to write down an algorithm as much simplified as possible,   
we shall replace the true potential with a carefully chosen model 
one: we use here a simple square barrier potential.
One has to imagine the two nuclei to be placed externally to the 
barrier, on the two opposite sides of it.   
The transmission factor for a particle coming, say, from the left 
with momentum asymptotically equal to $k_{l}$ is 
\begin{equation}
p_{c} = | { \exp[ -(i/2) L (k_l - 2 q + k_r)] 4 k_l  q \over 
          (k_l ((q + k_r) + \exp[2 i L q] (q - k_r)) + 
            q ((q + k_r) - \exp[2 i L q] (q - k_r))) }|^{2}
\end{equation}
The coefficients $q$, $k_{r}$ are the momenta respectively within the 
barrier and on the right side; $L$ is the barrier width. 
The momentum $k_{l}$ 
is obviously related to the momentum of the bound electron: if its 
binding energy is $ - E_{n}$, then an intuitive choice is to 
set $k_{l} = \sqrt{2 |E_{n}|}$. We choose to consider energy-conserving 
collisions, thus we set $k_{r} = k_{l}$.
In the original problem (Fig. 2), the energy needed by the 
electron to reach the top of the potential hill is $\Delta E = 
U_{M} - E_{n} = - |U_{M}| + |E_{n}|$.   
We define the height of the square potential barrier $V_{0}$ 
by keeping constant and equal to $\Delta E$ the energy deficit between 
potential and kinetic energy: this means
\begin{equation}
		V_{0} - {k_{l}^{2} \over 2} = \Delta E = - |U_{M}| + |E_{n}| \to
		V_{0} = 2 |E_{n}| +  {1 \over R} \left(\sqrt{Z_{t}} + 
	\sqrt{Z_{p}}\right)^{2} 
\end{equation}
(where we have used Eq. \ref{eq:um}). This relation defines $q$.
Finally, it is clear that the barrier width 
$L$ must be related to the internuclear distance $R$: we wish 
to have a zero-potential region ($ V \approx 0$) close to either of the 
nuclei. Basing upon indetermination relations, an electron bound to 
one nucleus, with kinetic energy $k_{l}^{2}/2$, moves within a region 
of spatial extent $\Delta \approx 1/k_{l}$. We choose this as the 
width of the potential-free region and set therefore $L = R - 2 \Delta$. Of course,
one must also set $L = 0 $ when $R < 2 \Delta$. Some tests showed that only 
minor differences are found if $L$ is allowed to vary slightly. For 
example, results shown in the next section remain almost unvaried by using the simpler 
choice $L = R$.  \\
The problem is, at this stage, reduced to performing a double 
integration: one over time for computing $P(b)$, and the other one over $b$ for getting 
$\sigma$. Neither of the two quadratures can be done analitically; 
however, they can be performed rather easily by using 
any standard mathematical software package.

\section{Numerical results}
\begin{figure}
	\includegraphics[scale=0.6]{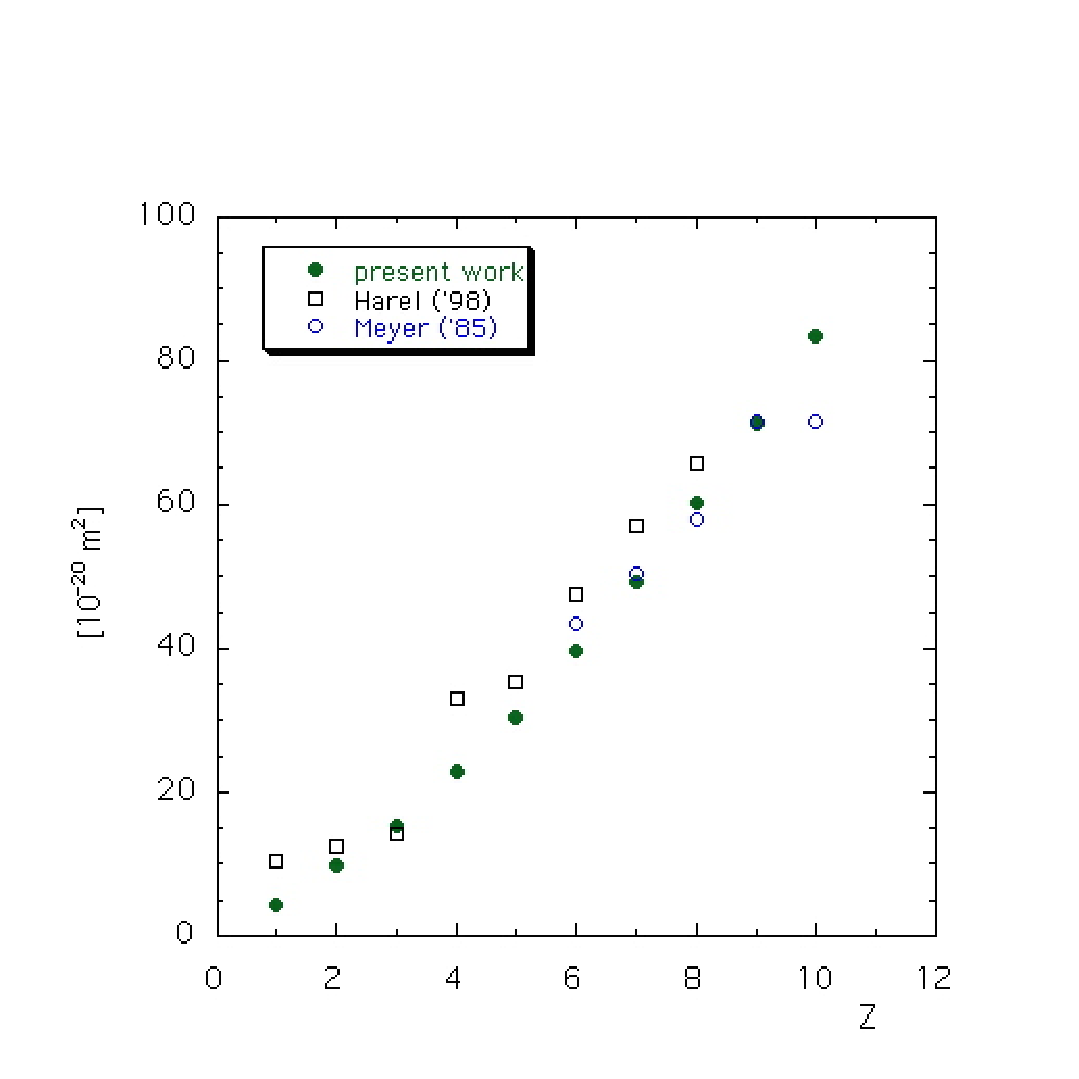}
\caption{\label{fig:risultati} Charge exchange cross section versus projectile charge.
Open circles, data from ref. \cite{meyer}; 
open squares, data from ref. \cite{harel}; full circles, present model. }
\end{figure}

We benchmark the model against experimental results from ref. 
\cite{meyer} and the theoretical ones coming from the molecular 
approach simulation of ref. \cite{harel}. In Fig. (\ref{fig:risultati}) we 
show some typical results for impacts between multicharged hydrogenlike ions and ground state hydrogen. 
In all cases, impact velocity is about 1/2 a.u. (it is exactly this value for 
the numerical results, while in experiments $ 0.49 \leq u \leq 0.51 $). 
The agreement is fairly good, with our model yielding a slight 
underestimate of theoretical results, but the accordance with 
experiment is pretty nice. The only exception is the $Z = 10$ case 
but, there, it is probable that is the experimental value to be flawed, 
since it departs rather abruptly from the general trend. \\
Tests carried on also for different velocities yielded results of comparable accuracy, 
with some caveat: see next section. 

\section{Discussion}
Besides being a very simple model to implement, and still being 
apparently rather accurate, a remarkable feature of this model is that it is self-consistent: 
although the parameters $L, V_{0}, k_{l}, \ldots$, have been guessed 
on the basis of order-of-magnitude reasonings, none of them is left to the experiment.  
However, the choice of the model potential has been arbitrary, 
constrained only by the condition that it must provide analytical 
expressions for $p_c$. One could, therefore, wonder if even better results can be got from 
a different choice of the model potential. 
The only other such potential we are aware of is the Eckart potential \cite{eckart}:
it is an approximately bell-shaped potential, and therefore rather different 
from the curve of Fig. 2. It has, however, the advantage 
of being smooth, without unphysical discontinuities. We have done a 
few tests using it: on the whole, we did find-not surprisingly-a worsening of the 
performances of the algorithm. \\
A discussion is, of course, necessary about the range of validity of 
the algorithm.
First of all, care must  be taken when trying to apply this model to different velocity 
regimes: at very low velocity $u << 1$ the straight-line approximation for nuclear motion fails. 
More important, according to a Feynman-like picture, 
the electron has more time to ``sample'' non-rectilinear paths 
connecting the two nuclei, thus making less correct the reasoning 
that here yielded to estimate $p_{c}$. For high-$u$ (say, $u \ge 1$), instead, ionization becomes relevant and 
the adiabatic hypothesis breaks down.  \\
Besides the impact velocity, $\sigma$ has a functional dependence 
upon a number of other parameters, e.g., projectile charge $Z_{p}$. It is 
straightforward to recover from Fig. 3 a power-law behaviour for 
this parameter: $\sigma $ is well fitted by a second-order polynomial. 
Roughly speaking, a $Z_{p}^{1}$ contribution comes from $p_{c}$, and 
another one from the effective range of interaction. This functional dependence 
is stronger that that usually quoted in literature (which is closer to 
$Z_{p}^{1}$) \cite{janev}. This could cause some trouble when trying 
to study highly-charged-ion collisions.


\begin{references}

		%% referenze secondo PR
%% 
 % \bibitem{rakovic} M.J. Rakovi\'c, D.R. Schultz, P.C. Stancil and R.K. 
 % Janev, J. Phys. A: Math. Gen. {\bf 34}, 4753 (2001)
 % 
 % \bibitem{schultz} D.R. Schultz, P.C. Stancil and M.J. Rakovi\'c, J. 
 % Phys. B: At. Mol. Opt. Phys. {\bf 34}, 2739 (2001)
 % 
 % \bibitem{ryufuku} H. Ryufuku, K. Sasaki and T. Watanabe, Phys. Rev. 
 % A, {\bf 21}, 745 (1980) 
 % 
 % \bibitem{niehaus} A. Niehaus, J. Phys. B: At. Mol. Phys. {\bf 19}, 
 % 2925 (1986)  
 % 
 % \bibitem{ostrovsky} V.N. Ostrovsky, J. Phys. B: At. Mol. Opt. Phys.
 % {\bf 28}, 3901 (1995) 
 % 
 % \bibitem{jpb} F. Sattin, J. Phys. B: At. Mol. Opt. Phys. {\bf 33} 
 % 861, 2377 (2000)
 % 
 % \bibitem{pra00} F. Sattin, Phys. Rev. A {\bf 62}, 042711 (2000) 
 % 
 % \bibitem{pra01} F. Sattin, Phys. Rev. A {\bf 64}, 034704 (2001) 
 % 
 % \bibitem{caillar} J. Caillar, A. Dubois, and J. P. Hansen, J. Phys. B: 
 % At. Mol. Opt. Phys. {\bf 33}, L715 (2000)
 % 
 % \bibitem{meyer} F.W. Meyer, A.M. Howald, C.C. Havener and R.A. Phaneuf, 
 % Phys. Rev. A {\bf 32}, 3310 (1985)
 % 
 % \bibitem{harel} C. Harel, H. Jouin and B. Pons B, At. Data Nucl. Data 
 % Tables {\bf 68}, 279 (1998)
 % 
 % \bibitem{eckart} C. Eckart, Phys. Rev. {\bf 35}, 1303 (1930)
 % 
 % \bibitem{janev} R.K. Janev, Phys. Lett. A {\bf 160}, 67 (1991)
 %%


		%% referenze secondo PLA
		 % 
\bibitem{rakovic} M.J. Rakovi\'c, D.R. Schultz, P.C. Stancil and R.K. 
Janev, J. Phys. A: Math. Gen. 34 (2001) 4753.

\bibitem{schultz} D.R. Schultz, P.C. Stancil and M.J. Rakovi\'c, J. 
Phys. B: At. Mol. Opt. Phys. 34 (2001) 2739.

\bibitem{ryufuku} H. Ryufuku, K. Sasaki and T. Watanabe, Phys. Rev. 
A 21 (1980) 745. 

\bibitem{niehaus} A. Niehaus, J. Phys. B: At. Mol. Phys. 19 (1986) 
2925.  

\bibitem{ostrovsky} V.N. Ostrovsky, J. Phys. B: At. Mol. Opt. Phys.
28 (1995) 3901. 

\bibitem{jpb} F. Sattin, J. Phys. B: At. Mol. Opt. Phys. 33 
(2000) 861, 2377.

\bibitem{pra00} F. Sattin, Phys. Rev. A 62 (2000) 042711. 

\bibitem{pra01} F. Sattin, Phys. Rev. A 64 (2001) 034704. 

\bibitem{caillar} J. Caillar, A. Dubois, and J. P. Hansen, J. Phys. B: 
At. Mol. Opt. Phys. 33 (2000) L715.

\bibitem{meyer} F.W. Meyer, A.M. Howald, C.C. Havener and R.A. Phaneuf, 
Phys. Rev. A 32 (1985) 3310.

\bibitem{harel} C. Harel, H. Jouin and B. Pons B, At. Data Nucl. Data 
Tables 68 (1998) 279.

\bibitem{eckart} C. Eckart, Phys. Rev. 35 (1930) 1303.

\bibitem{janev} R.K. Janev, Phys. Lett. A 160 (1991) 67.


\end{references}
\end{document}